%% file: main.tex
\documentclass[acmtog]{acmart}
\input{text/0_preamble}

\begin{document}

\title{Learned Localized Mesh Refinement}

\author{Xiao Zhan}
\orcid{0000-0002-1825-0097}
\affiliation{
  \institution{Massachusetts Institute of Technology (MIT)}
  \city{Cambridge}
  \state{Massachusetts}
  \country{USA}
}
\email{zhanx@mit.edu}

\author{Chrystiano Ara\'{u}jo}
\orcid{0000-0002-8237-9426}
\affiliation{
  \institution{Roblox}
  \city{San Mateo}
  \state{California}
  \country{USA}
}
\email{caraujo@roblox.com}

\author{Kangle Deng}
\orcid{0009-0000-0565-4255}
\affiliation{
  \institution{Roblox}
  \city{San Mateo}
  \state{California}
  \country{USA}
}
\email{kdeng@roblox.com}

\author{Maneesh Agrawala}
\orcid{0000-0002-8996-7327}
\affiliation{
  \institution{Stanford}
  \city{Stanford}
  \state{California}
  \country{USA}
}
\email{maneesh@cs.stanford.edu}

\author{Hsueh-Ti Derek Liu}
\orcid{0009-0001-1753-4485}
\affiliation{
  \institution{Roblox}
  \city{Vancouver}
  \state{British Columbia}
  \country{Canada}
}
\email{hsuehtiliu@roblox.com}

\author{Mina Konakovi\'{c} Lukovi\'{c}}
\orcid{0000-0002-2895-0206}
\affiliation{
  \institution{Massachusetts Institute of Technology (MIT)}
  \city{Cambridge}
  \state{Massachusetts}
  \country{USA}
}
\email{minakl@mit.edu}

% \renewcommand{\shortauthors}{Anonymous Author(s)}

\input{text/0_abstract}






\begin{teaserfigure}
\centering
\includegraphics[width=\textwidth]{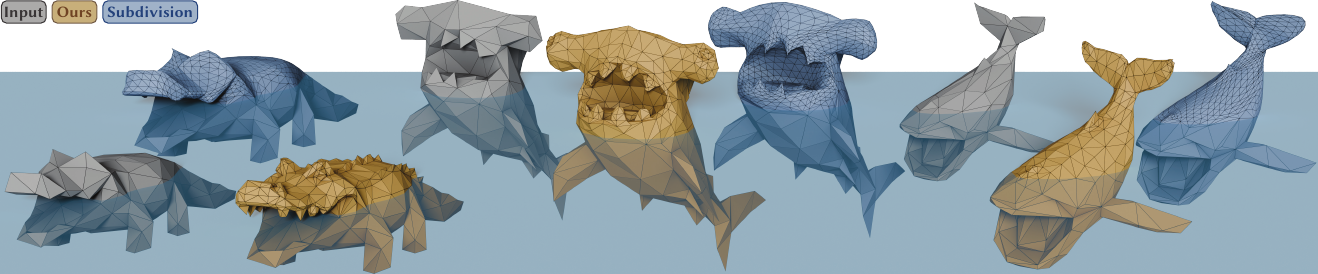}
\caption{
Refining aquatic animals as they breach.
Our neural model (yellow) locally refines coarse meshes (gray) by adding shape-aware geometric details (alligator, shark), which cannot be recovered by classical butterfly subdivision (blue), as it smooths inputs toward their limit surfaces (acceptable for whale). 
}
\Description{}
\label{fig:refinement_against_classical}
\end{teaserfigure}

\maketitle

\input{text/1_intro}
\input{text/2_0_related}
\input{text/2_1_background}
\input{text/3_overview}

\input{text/4_progressive}

\input{text/5_gen_progressive}

\input{text/6_results}
\input{text/7_conclusion}

\begin{acks}
We thank Victor Zordan, Ahmed Mahmoud, and the MIT Algorithmic Design Group for insightful discussions during the project.
We thank reviewers for their valuable feedback.
Xiao Zhan acknowledges the generous support of the MathWorks Engineering Fellowship. 
The MIT Algorithmic Design Group acknowledges the generous support of the Wistron Corporation and the JPMorgan Chase-CSAIL Research Award.
\end{acks}

% \clearpage

\bibliographystyle{ACM-Reference-Format}
\bibliography{bibliography}

% \clearpage

\input{text/8_supp}
\end{document}

%% file: text/0_preamble.tex
\copyrightyear{2026}
\acmYear{2026}
\setcopyright{cc}
\setcctype{by}
\acmConference[SA Conference Papers '26]{SIGGRAPH Asia 2026 Conference Papers}{December 01--04, 2026}{Kuala Lumpur, Malaysia}
\acmBooktitle{SIGGRAPH Asia 2026 Conference Papers (SA Conference Papers '26), December 01--04, 2026, Kuala Lumpur, Malaysia}
\acmDOI{10.1145/3829340.3842280}
\acmISBN{979-8-4007-2842-6/2026/12}

\usepackage{natbib}
\usepackage{subcaption}
\usepackage{mathtools}
\usepackage{xcolor}
\usepackage{wrapfig}

\usepackage[capitalize,noabbrev]{cleveref}

\usepackage{siunitx}
\usepackage{booktabs,multirow}  % for metrics
\setcitestyle{square}

\input{text/macros}

%% file: text/macros.tex
\definecolor{darkblue}{rgb}{0.0, 0.2, 0.6}

\usepackage{siunitx}
\newcommand{\numpct}[1]{\num[scientific-notation=fixed,fixed-exponent=-2,drop-exponent]{#1}}
\newcommand{\nummmd}[1]{\num[scientific-notation=fixed,fixed-exponent=-3,drop-exponent]{#1}}
\newcommand{\numjsd}[1]{\num[scientific-notation=fixed,fixed-exponent=-1,drop-exponent]{#1}}
\newcommand{\numemdq}[1]{\num[scientific-notation=fixed,fixed-exponent=-2,drop-exponent]{#1}}
\newcommand{\bnumpct}[1]{{\boldmath\textbf{\numpct{#1}}}}
\newcommand{\bnummmd}[1]{{\boldmath\textbf{\nummmd{#1}}}}
\newcommand{\bnumjsd}[1]{{\boldmath\textbf{\numjsd{#1}}}}
\newcommand{\bnumemdq}[1]{{\boldmath\textbf{\numemdq{#1}}}}

\newcommand{\numfourk}[1]{\num[round-mode=figures, round-precision=2]{\fpeval{#1 * 1000}}}

\newcommand{\myboxed}[2][black!10]{%
  \setlength{\fboxrule}{0pt} \setlength{\fboxsep}{1pt}%
  \rlap{\hspace*{\fboxsep}%
    \phantom{$\displaystyle#2$}}%
    \smash{\colorbox{#1}{$\displaystyle#2$}}}

\newcommand*{\refsec}[1]{%
  \begingroup
    \def\sectionautorefname{Section}%
    \def\subsectionautorefname{Section}%
    \def\subsubsectionautorefname{Section}%
    \autoref{sec:#1}%
  \endgroup
}
\newcommand*{\refequ}[1]{%
  \begingroup
    \def\equationautorefname{Eq.}
    \autoref{eq:#1}%
  \endgroup
}

\newcommand*{\reffig}[1]{%
  \begingroup
    \def\figureautorefname{Fig.}%
    \autoref{fig:#1}%
  \endgroup
}

\newcommand*{\reftab}[1]{%
  \begingroup
    \def\tableautorefname{Tab.}%
    \autoref{tab:#1}%
  \endgroup
}
\newcommand*{\refapp}[1]{%
  \begingroup
    \def\appendixautorefname{App.}%
    \autoref{app:#1}%
  \endgroup
}

\newcommand{\R}{\mathbb{R}}
\newcommand{\M}{\mathcal{M}}
\newcommand{\N}{\mathcal{N}}
\newcommand{\F}{\mathcal{F}}

\newcommand{\vecFont}[1]{\mathbf{#1}}

\def\vp{{\vecFont{p}}}

%% file: text/0_abstract.tex
\begin{abstract}
We present a neural method for adaptive triangle mesh refinement, in which an autoregressive model adds geometric detail to selected regions of an input mesh while leaving the rest unchanged, a key capability for efficiently allocating mesh budget.
Existing upsampling methods struggle to achieve this. Classical subdivision schemes refine triangulation without semantic awareness of the underlying shape or the ability to recover geometric details missing from a coarse input. Recent neural mesh models generate shapes globally, sacrificing region-specific control.
We propose a novel tokenizer that yields combinatorially many valid upsampling trajectories from a single mesh.
Trained on such data, our locally-conditioned autoregressive architecture allows for direct manipulation of topology and geometry within target regions of an input mesh. 
We validate our method against state-of-the-art approaches and demonstrate its ability to perform adaptive upsampling with region-selective control, a capability absent from existing approaches.
This unlocks inference-time view-dependent refinement, physics-aware region refinement, and coarse-shape conditioned novel mesh synthesis.
We provide code at \href{https://github.com/seanxzhan/learned-localized-mesh-refinement/}{\textcolor{darkblue}{github.com/seanxzhan/learned-localized-mesh-refinement}}. 
\end{abstract}

%% file: text/1_intro.tex
\section{Introduction}

Refining triangle meshes, the operation of subdividing faces while relocating vertices, is a fundamental subroutine in 3D content creation, simulation, and rendering. 
Artists iterate on proxy geometries to refine them into production-quality assets, engineers refine meshes when physical accuracy demands it, and real-time systems (such as video games) stream incremental details under tight resource budgets.
In such cases, the operations call for \textit{adaptively} refining specific mesh regions while keeping the rest of the shape fixed.

Neural approaches offer a promising solution. Learned data-driven priors can refine geometric details that adhere to shape semantics, rather than relying on classical subdivision schemes that smooth shapes toward their limit surface regardless of semantic structure. While recent autoregressive models directly generate meshes~\cite{meshgpt, meshanything-v2, wang2026face, zhao2026lato}, progressive-mesh variants~\cite{vertexregen, armesh} extend to coarse-to-fine generation. 
However, these methods autoregressively synthesize the mesh as a single ordered sequence of upsampling steps, with no way to refine only a specific region.
The limitation lies in the tokenization: where details appear is determined by the network's learned canonical traversal over the entire mesh rather than any external signal that specifies a region of interest.

In this work, we take inspiration from progressive meshes~\cite{progressivemeshes, Hoppe97} and introduce a learning-based solution for \emph{adaptive mesh refinement}.
The core of our method is a novel dependency-free tokenizer that organizes refinement operations into a binary forest of vertex-split operations.
Since vertex splits in different subtrees are independent, any vertex present in the current mesh can be split next.
In this way, a single mesh can be sampled to yield combinatorially many valid token sequences, providing a rich training signal across arbitrary upsampling orders.
We additionally propose an autoregressive architecture conditioned on local geometry at each upsampling step, learning shape priors anchored to spatial regions rather than to global sequence positions.

We evaluate our method against two state-of-the-art progressive-mesh autoregressive models~\cite{vertexregen, armesh} and show that our tokenization enable region-selective refinements that baselines cannot achieve (\reffig{results_compare_adap_inf}).
Additionally, our model produces more accurate refinements of unseen coarse inputs than baselines, both quantitatively (\reftab{results_metrics}) and qualitatively (\reffig{results_recon_refine}). 
Our novel capability of adaptive upsampling unlocks view-dependent refinement at inference time to satisfy face count constraints (\reffig{results_adaptive_inference}), as well as coarse-mesh conditioned novel shape synthesis (\reffig{results_editing_deer}, \reffig{results_editing_mix_and_match}).
In contrast to traditional subdivision methods, our data-driven method produces locally refined geometry that adheres to shape semantics (\reffig{refinement_against_classical}), or more closely matches target simulation (\reffig{results_shiphulls}). 

Our contributions include a novel dependency-free tokenizer, a conditional autoregressive model, and the pipeline for learning adaptive mesh refinement. 

%% file: text/2_0_related.tex
\section{Related Work}\label{sec:related_work}

\paragraph{Adaptive Level-of-Detail}

\begin{wrapfigure}[6]{r}{1.4in}
    \vspace{-10pt}
	\includegraphics[width=\linewidth, trim={5mm 0mm 0mm 0mm}]{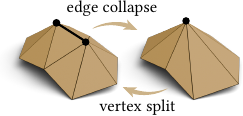}
	\label{fig:edge_collapse_vertex_split} 
\end{wrapfigure}
Mesh level of detail (LOD) techniques reduce mesh resolution while maintaining its visual appearance \cite{luebke2002level}. 
% 
% An approach closely related to ours is the \emph{progressive meshes} introduced by \citet{progressivemeshes}, which encodes a mesh at all resolutions where consecutive levels differ by a single edge collapse, or equivalently by its inverse, a vertex split.
An approach closely related to ours is the \emph{progressive meshes} introduced by \citet{progressivemeshes}, which encodes a mesh at all resolutions where consecutive levels differ by a single edge collapse.
It enables granular control over mesh resolution and motivates the development of \emph{adaptive refinement}: upsampling a local patch while holding other regions of a mesh fixed.
At the core of adaptive refinement is the design of preconditions between vertex splits so that one can correctly update the local mesh connectivity.
Different preconditions are based on the existence of a subset of one-ring vertices \cite{progressivemeshes, XiaV96}, edges \cite{Pajarola01, PajarolaD04}, or faces \cite{Hoppe97}. 
Although they can achieve adaptive refinement, too many preconditions lead to long dependency chain effect: refining a region requires refining other dependent regions first, leading to unnecessary computations and random memory access. 
This motivates approaches such as \cite{Ponchio2009} to restrict refinement to small patches with fixed boundary vertices, reducing dependency chains at the cost of lower simplification quality.
In lieu of this, \citet{KimL01} propose to infer connectivity changes from the connectivity of vertex duals, eliminating most preconditions between vertex splits.
When combined with data structures that support non-manifold meshes \cite{DerzapfG12}, this approach leads to \emph{dependency-free} progressive meshes that supports vertex splits at any arbitrary order, as long as the vertex to be split exists in the current mesh.
Progressive meshes literature set up the foundation of how to adaptively refine a given triangle mesh. However, they are not applicable to mesh generation because they require the knowledge of the high-resolution input mesh, which is not available during inference. 
We thus propose novel tokenization and training strategies, enabling autoregressive adaptive mesh generation.

\paragraph{Mesh Refinement and Upsampling}
\emph{Subdivision surfaces} is a popular technique for refining coarse geometries to denser ones, in the absence of high-resolution targets.
The core idea relies on defining a smooth surface as the limit of a sequence of recursive mesh refinement steps. This has inspired a wide variety of subdivision schemes based on different families of limit surfaces, such as \cite{loop1987smooth, catmull1998recursively, butterflysubdivision-dyn, butterflysubdivision-zorin}. For a comprehensive survey of subdivision, we refer the reader to \cite{zorin2000subdivision}.
While subdivision is highly effective at creating smooth geometries, the property of converging to a smooth limit surface prevents them from producing complex geometric details (\reffig{refinement_against_classical}).
This motivates explorations on how to incorporate data-driven priors into subdivision schemes such that the limit surface is learned from a collection of shapes.
Neural Subdivision~\cite{neuralsubdivision} relies on a neural network to predicts vertex offsets at each subdivision iteration conditioned on the local surface patch. 
Neural Progressive Meshes~\cite{neuralprogressivemeshes} uses subdivision-based encoder-decoder to learn surface patterns from a collection of shapes and progressively transmits higher resolution features.
Wile they successfully inject data-driven priors, their reliance on purely local geometric features limits their ability to produce structured details that require global knowledge.
Other approaches have explored how to refine geometries and hallucinate structured details. They focus on refining geometries represented as voxels~\cite{decorgan, decollage, artdeco}, displacement maps~\cite{genvdm}, and implicit functions~\cite{magicclay, texturegeometryrefinement}.
Although these works show promise in adding geometric details, as a mesh processing tool, they are destructive: they break the carefully crafted mesh topology during intermediate steps, such as iso-surface extraction from a refined implicit function.
Our work complements these approaches by directly operating at the triangle mesh level, making localized changes only within the region of interest. In contrast to subdivision-based techniques, our method generates semantic details where standard subdivision fails (\reffig{refinement_against_classical}, \reffig{results_shiphulls}).

\paragraph{Autoregressive Mesh Generation}
To explicitly model mesh topology, recent approaches formulate shape generation as an autoregressive sequence learning problem.
Early works like PolyGen~\cite{polygen} decompose the task into autoregressively predicting vertices followed by face connectivity. 
Subsequent methods~\cite{meshgpt, meshanything} train autoregressive models in the latent space of discrete face sequences.
Additional works~\cite{chen2024meshxl, zhao2026lato} have explore novel neural representations for mesh learning.
To scale to higher resolutions, other efforts~\cite{meshanything-v2, edgerunner, meshtron, wang2026face, deepmesh} have explored compact tokenization strategies. 

Despite improved token efficiency, previous methods fundamentally rely on face-by-face generation. Consequently, their intermediate outputs are topologically incomplete meshes that are unavailable for downstream use until the entire sequence is generated. 
In contrast, progressive generation methods operate on coarse-to-fine sequences, ensuring a coherent, complete mesh at ever step.
VertexRegen~\cite{vertexregen} adopts the progressive mesh representation, formulating generation as a 1D sequence of vertex split operations where the model only predicts vertex positions to update geometry and connectivity. 
Similarly, ARMesh~\cite{armesh} operates on progressive simplicial complexes~\cite{progressivesimplicialcomplexes}, allowing for upsampling from a single vertex.
While utilizing vertex split operators similar to ours, these methods focus primarily on global shape generation from scratch, lacking mechanisms for fine-grained local control. 
Our approach enables localized, adaptive refinement by training an autoregressive model on diverse refinement sequcenes sampled with our novel tokenizer, where we explicitly condition the model on vertex one-ring neighbors. 
This design supports downstream applications such as view-dependent inference (\reffig{results_adaptive_inference}), physics-driven local upsampling (\reffig{results_shiphulls}), and coarse-shape conditioned novel shape synethsis (\reffig{results_editing_deer}, \reffig{results_editing_mix_and_match}).

%% file: text/2_1_background.tex
\section{Background}\label{sec:background}

\begin{figure}[t!]
\centering
\includegraphics[width=\linewidth]{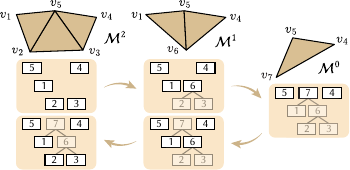}
\caption{
Recording edge collapses (top) from the fine mesh ($\mathcal{M}_2$) to the coarsest level ($\mathcal{M}_0$) in a binary forest (bottom) and using it to perform vertex splits. 
When $v_2, v_3$ is collapsed to $v_6$, a parent node $6$ is created and child nodes $2, 3$ become \textit{inactive} (faded color). Collapsing into $v_7$ keeps the history of all past collapses.
Conversely, when $v_7$ is split into $v_1, v_6$, the node 7 becomes inactive and $1, 6$ become \textit{active} (full color).
}
\Description{}
\label{fig:background_binary_forest}
\end{figure}

Progressive meshes, introduced by \citet{progressivemeshes}, is a scheme to store and transmit a 3D object at different resolutions.
Traditionally, a progressive mesh has been represented either as a sequence or as a directed acyclic graph of vertex split operations that can be performed on a coarse mesh to add detail and refine it.
These representations track both direct parent-child dependencies and \emph{indirect} dependencies between vertex splits that are distant in the edge collapse sequence. 
However, as pointed out by Kim and Lee\,\shortcite{KimL01} and Derzapf et al.\,\shortcite{DerzapfG12}, indirect dependencies can be eliminated and the progressive mesh can be represented as a \emph{binary forest}.
This binary forest representation $\F$ only contains direct dependencies.

Given a triangle mesh $\M^L$, a progressive mesh starts by constructing its lower resolution counterparts with a series of edge collapse $c_l$ operations
\begin{align}
\label{eq:pm_construction}
    \M^L \xrightleftharpoons[s_{L-1}]{c_{L-1}} \M^{L-1} \xrightleftharpoons[s_{L-2}]{c_{L-2}} \cdots \xrightleftharpoons[s_{l}]{c_{l}} \M^{l} \xrightleftharpoons[s_{l-1}]{c_{l-1}}\cdots\xrightleftharpoons[s_{1}]{c_{1}} \M^{1} \xrightleftharpoons[s_{0}]{c_{0}} \M^{0}. 
\end{align}
Each subsequent level $\M^{l}$ is computed by collapsing a single edge $(i,j)$ from its previous level $\M^{l+1}$, and forming a new vertex $k$. For each edge collapse, the progressive mesh stores information about geometry and connectivity changes (see \cite{KimL01} for more details), so that the process can \emph{be losslessly} reversed by performing a vertex split $s_l$ on the vertex $k$ to recover the edge $(i,j)$.
A binary forest $\F$ can be built bottom-up from these edge collapse operations $\{ c_l \}$ (\reffig{background_binary_forest}). Specifically, if an edge $(i, j)$ collapses into the vertex $k$, then in the binary forest, the vertex $k$ becomes the parent node of the child nodes $i,j$. We then mark $i, j$ \textit{inactive} (faded color) and $k$ as \textit{active} (full color). 

During the reverse process $\{s_l\}$, a vertex $k$ splits into an edge $(i, j)$ by \textit{deactivating} the node $k$ and \textit{activating} $i, j$. 
An important observation is that the leaf nodes in the trees correspond to the vertices of the input mesh $\M^L$, and the root nodes correspond to the vertices in the coarsest mesh $\M^0$.
We define an \textit{active ancestor} as the active node that represents the vertex at a current mesh level $\M^l$ that is an ancestor of a vertex at $\M^L$.
For example, in the rightmost forest in \reffig{background_binary_forest}, node 7 is an active ancestor of the vertex $v_3$.

\begin{figure}[t!]
\centering
\includegraphics[width=\linewidth]{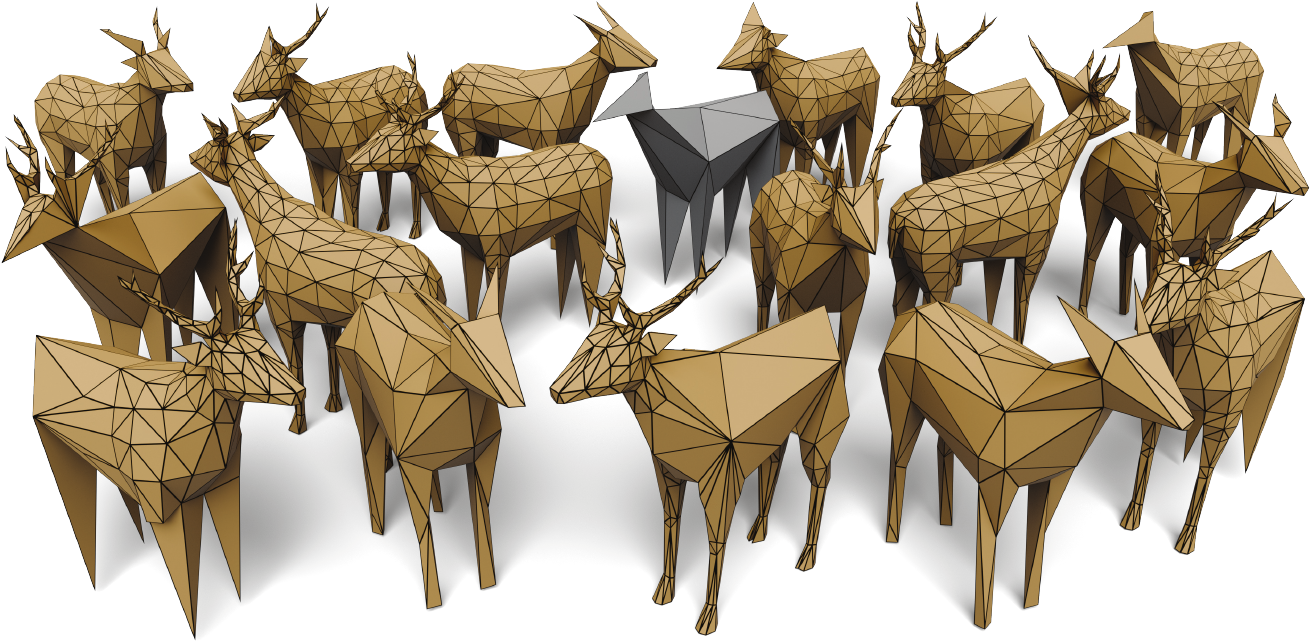}
\caption{
From a single coarse mesh (gray), our dependency-free tokenizer samples diverse triangulations, each with distinct connectivity and geometry, providing rich contexts for our locally-conditioned autoregressive model.
}
\Description{}
\label{fig:triangulations}
\end{figure}

%% file: text/3_overview.tex
\section{Method Overview} \label{sec:approach}
Our method enables learning-based local mesh refinement. 
We convert each training mesh into a progressive mesh representation that explicitly exposes vertex split operations, and we introduce a dependency-free tokenizer that encodes these operations without requiring a fixed refinement order (\refsec{arbitrary_upsampling}).
Our formulation enables sampling diverse sequences for training an autoregressive model.
To locally refine a target area from a coarse mesh, we condition the autoregressive model on the geometric features of each vertex's one-ring neighborhood (\refsec{gen_progressive}).
We evaluate our framework against state-of-the-art progressive refinement methods and demonstrate view-dependent refinement, physics-driven local upsampling, and coarse-mesh conditioned shape synthesis (\refsec{results}).

%% file: text/4_progressive.tex
\section{Diversified Upsampling via Dependency-Free Tokenizer} \label{sec:arbitrary_upsampling}

\begin{figure}[t!]
\centering
\includegraphics[width=\linewidth]{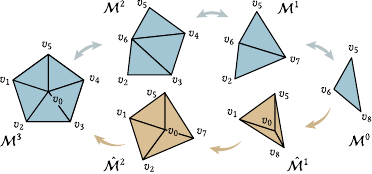}
\caption{
The local neighborhood of a vertex may change for different upsampling sequences. When reversing the edge collapse order by splitting $v_8$ from $\M^0$ to $\M^1$ (blue), the resulting vertices $v_2, v_7$ share a face with $v_6$. However, if $v_8$ is split from $\hat{\M}^1$ to $\hat{\M}^2$ (yellow), $v_2, v_7$ share a face with $v_0$. 
}
\Description{}
\label{fig:method_arb_up}
\end{figure}

Our insight is that shuffling the vertex split order produces multiple refinement sequences per mesh, exposing diverse upsampling triangulations (\reffig{triangulations}).
We propose a novel \emph{dependency-free} tokenizer to sample such sequences to train our autoregressive model.

We first introduce \textit{vertex split tokens} that encode splitting a collapsed vertex at level $\M^l$ to recover $\M^{l+1}$, including the geometry (\refsec{geometry_token}) and topology (\refsec{vertex_split_tokens}) components.
We then detail how to encode these split tokens such that they support out-of-order splits on an arbitrary vertex at level $\M^l$ (Section~\ref{sec:topology_tokenizer}), rather than requiring the exact reverse order of the edge collapse sequence in Eq.~\ref{eq:pm_construction}. 
By splitting an arbitrary vertex, we arrive at a new mesh $\hat{\M}^{l+1}$ that has the same vertex count but with a different triangulation from $\M^{l+1}$ in the edge collapse sequence (\reffig{method_arb_up}), hence augmenting the training set and exposing the autoregressive model to a variety of triangulations at each resolution level.
Our method ensures lossless recovery of the input $\M^L = \hat{\M}^L$ at the end of refinement, while producing diverse refinement sequences.

%%%%%%%%%%%%%%%%%%%%%%%%%%%%%%%%%%%%%%%%%%%%
% Vertex Split Tokens
%%%%%%%%%%%%%%%%%%%%%%%%%%%%%%%%%%%%%%%%%%%%

\subsection{Vertex Split Geometry Tokens} \label{sec:geometry_token}
Splitting vertex $k$ into $i, j$ relocates vertices.
Let $\vp_i, \vp_j, \vp_k \in \R^3$ be vertex positions. The \emph{geometry} tokens for vertex $k$ are
\begin{align}\label{eq:geometry_token}
    \gamma_k = \{ \vp_{ki}, \vp_{ji}\},
\end{align}
where $\vp_{ki} = \vp_i - \vp_k$ denotes the displacement between vertex $k$ and $i$.
We refer readers to App. A on quantization details.

\subsection{Vertex Split Topology Tokens} \label{sec:vertex_split_tokens}

\begin{figure}[t!]
\centering
\includegraphics[width=\linewidth]{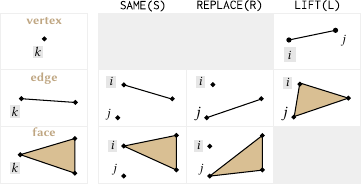}
\caption{
Enumeration of all \textit{vertex split token} assignments. 
To split a vertex $\myboxed{k}$, there are three cases $\{\texttt{S}, \texttt{R}, \texttt{T}\}$ to update $\myboxed{k}$'s one-ring neighbors (left column) to $\myboxed{i}, j$ (right table). Note that the gray background under $\myboxed{k}, \myboxed{i}$ denotes that any neighbor that connects with $\myboxed{i}$ updates its connectivity to $\myboxed{k}$; whether $i$ or $j$ exhibits this behavior is an implementation choice. 
}
\Description{}
\label{fig:method_split_tokens_table}
\end{figure}

\begin{figure}[t!]
\centering
\includegraphics[width=\linewidth]{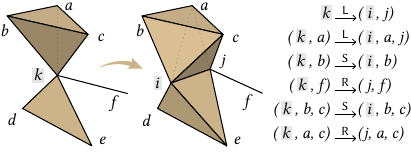}
\caption{
An example of decoding a vertex split that covers all token cases.
Our vertex split tokens are compatible with non-manifold meshes. 
}
\Description{}
\label{fig:method_split_tokens_worked_example}
\end{figure}

In addition to geometry changes, a vertex split updates the local topology.
Given two consecutive meshes $\M^{l+1}$, $\M^{l}$ for a single edge collapse in the progressive meshes sequence, we use 
\begin{align*}
    \N^l_k &\coloneqq \textit{star}\,({k}) \subseteq \M^l \\
    \N^{l+1}_{i, j} &\coloneqq \textit{star}\,(i) \cup \textit{star}\,({j}) \subseteq \M^{l+1},
\end{align*}
to denote the vertex one-ring neighbors $\N^l_k$ on $\M^l$ and the edge one-ring neighbors $\N^{l+1}_{i, j}$ on $\M^{l+1}$. The \textit{star} operation collects a set of 1-ring neighborhood simplices, including vertices, edges, and faces.
A vertex split transforms all the simplices from $\N^l_k$ to the simplices in $\N^{l+1}_{i, j}$. The transformation of each type of simplex can be compactly written using three operations   
\begin{align*}
\label{eq:split_tokens}
    \sigma_x \in \{\texttt{SAME}\,(\texttt{S}), \texttt{REPLACE}\,(\texttt{R}), \texttt{LIFT}\,(\texttt{L})\}.
\end{align*}
as illustrated in \reffig{method_split_tokens_table}, and we provide an example vertex split that covers all token cases in \reffig{method_split_tokens_worked_example}.
% When collapsing a single edge from $\M^{l+1}$ to $\M^{l}$, given a simplex $y \in \N^{l+1}_{i, j}$ and its corresponding simplex $x \in \N^l_k$, we use $\textit{Enc}_x: y \mapsto \sigma_x, \ \textit{Dec}_x: \sigma_x \mapsto y$
% %
% to denote the encoding process \textit{Enc} from mesh updates to our split tokens, and \textit{Dec} for the reverse.
The \texttt{SAME} and \texttt{REPLACE} operations keep the same simplex dimension after splitting (e.g., an edge remains an edge). The \texttt{LIFT} operation lifts a simplex to one dimension higher (e.g., an edge becomes a triangle). 
Some operations are undefined depending on the type of input simplex. For instance, \texttt{LIFT} is inapplicable to a 2-simplex (triangle) as raising up one dimension higher yields a 3-simplex (tetrahedron), which is above the highest simplex dimension considered in our setting.
We refer to these split operations $\sigma_x$ as \emph{split tokens}, which is the central ingredient to train an autoregressive model for localized mesh refinement.

When collapsing a single edge from $\M^{l+1}$ to $\M^{l}$, given a simplex $y \in \N^{l+1}_{i, j}$ and its corresponding post-collapse simplex $x \in \N^l_k$, the pair $(x,y)$ falls into exactly one of the cases
enumerated in \reffig{method_split_tokens_table}.
We write $\textit{Enc}_x: y \mapsto \sigma_x$ for reading off the token of that case and $\textit{Dec}_x: \sigma_x \mapsto y$ for the reconstruction of $y$.
Both functions are table lookups determined by the pair $(x,y)$ and the convention fixing which of $i,j$ inherits $k$'s connectivity.

\begin{figure*}
\centering
\includegraphics[width=\linewidth]{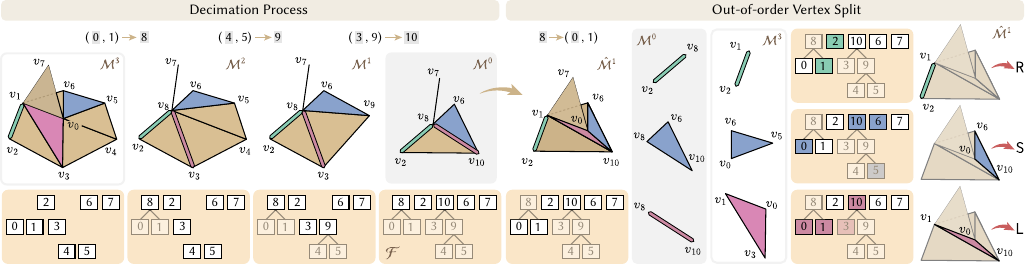}
\caption{
The process to produce topology tokens with our tokenizer from an out-of-order vertex split.
We visualize a sequence of edge collapses with the corresponding binary forest $\F$ constructed from the process on the left (``Decimation Process"). 
After edge collapses, we perform an out-of-order vertex split on $v_8 \to v_0, v_1$ from level $\M^0$ to $\hat{\M}^1$. 
The recorded tokens at the time of collapse $v_0, v_1 \to v_8$ no longer applies because they depend on $v_8$'s neighborhood at $\M^2$. 
Thus, we use our dependency-free tokenizer to obtain applicable topology tokens in the following three steps: 
(1) For $v_8$'s neighboring simplex at $\M^0$, identify the corresponding simplex in the original mesh level $\M^3$ (from 6th to 7th column).
(2) Traversing $\F$ bottom-up to find the \emph{active ancestry} nodes in the updated mesh level $\hat{\M}^1$ (full-colored nodes in the 8th column).
(3) These active ancestors define the new simplices that should be added to $\hat{\M}^1$ and allow us to encode their difference from the starting simplices on $\M^0$ into topology tokens (red arrow).
For example,  face $(v_8, v_{10}, v_6) \in \M^0$ corresponds to $(v_0, v_5, v_6) \in \M^3$, which traverses to $(v_0, v_{10}, v_6) \in \hat{\M^1}$, and we obtain split token $\texttt{S} = \text{Enc}_{v_8, v_{10}, v_6}(v_0, v_{10}, v_6)$.
}
\Description{}
\label{fig:method_dependency_free_tokens}
\end{figure*}

Directly storing the indices of simplices in $\textit{star}\,(k)$ and their split tokens at each mesh level during edge collapses, following the approach of Lei et al.\,\shortcite{armesh}, constrains vertex splits to the strict reverse order.
This limitation arises because $\textit{star}\,(k)$ may change during decimation.
If we wish to split $k$ at a different mesh level to obtain a novel upsampling order, the originally encoded split tokens $\{ \sigma_x \}$ reference simplices that no longer exist or have been merged (\reffig{method_dependency_free_tokens}).
This motivates our \emph{dependency-free topology tokenizer}, which enables splitting vertices in an arbitrary order.

%%%%%%%%%%%%%%%%%%%%%%%%%%%%%%%%%%%%%%%%%%%%
% dependency-free Token Encoding
%%%%%%%%%%%%%%%%%%%%%%%%%%%%%%%%%%%%%%%%%%%%

\subsection{Dependency-Free Topology Tokenizer} \label{sec:topology_tokenizer}
Enabling out-of-order vertex splits requires a tokenizer that is flexible to produce different tokens depending on the specific configuration of the local mesh neighborhood.
The key idea is to derive tokens from the binary forest $\mathcal{F}$ induced by the entire edge collapse sequence (Section~\ref{sec:background}).
We leverage the insight that, in progressive meshes, for a simplex $x \in \hat{\N}^l_k$ neighboring a vertex $k$ at an arbitrary mesh level $\hat{\M}^l$, there exists a unique corresponding simplex $x^L$ in the original mesh $\M^L$. Splitting $k$ into $i,j$ turns $x$ into the simplex
\begin{equation}\label{eq:tracing}
    \hat{x} = \mathcal{T}_{\uparrow}(x^L, \hat{\M}^{l+1} \mid \F)
    \; \in \hat{\N}^{l+1}_{i, j} \subseteq \hat{\M}^{l+1},
\end{equation}
where $\mathcal{T}_{\uparrow}$ takes in leaf nodes that correspond to vertices of $x^L$ and traverses the forest $\F$ bottom-up until reaching their active ancestor nodes. 
Note that prior to traversal, the activation pattern of $\F$ is updated such that the node for $k$ is deactivated and those for $i, j$ activated. 
The function $\mathcal{T}_{\uparrow}$ resolves the
connectivity of $\hat{\M}^{l+1}$ without reference to how $\hat{\M}^l$ was reached, which frees the vertex split from the edge collapse order.

We thus define our \emph{dependency-free tokenizer} as
\begin{equation*}
    \Sigma_\F(x, \hat{\M}^l) \coloneqq \textit{Enc}_x(\hat{x}) = \sigma_x,
\end{equation*}
which assigns a split token to $x$ by matching the pair $(x, \hat{x})$ against the cases enumerated in \reffig{method_split_tokens_table} (\refsec{vertex_split_tokens}).
\reffig{method_dependency_free_tokens} illustrates the construction of the binary forest $\F$ during decimation, followed by an out-of-order vertex split using $\Sigma_\F$ to infer the split tokens. 

As noted by \citet{KimL01}, the formulation in \refequ{tracing} allows us to split vertices in nearly any order, subject to manifold constraints. If we further permit non-manifold intermediate configurations (as in \cite{DerzapfG12}), $\Sigma_\F$ enables splitting vertices in any order without the need of handling \emph{indirect} dependencies. We therefore refer to our tokenizer as a \textit{dependency-free} vertex split tokenizer.
Moreover, since our tokens are defined over simplicial complexes (\reffig{method_split_tokens_table}) rather than manifold surfaces, the input mesh may itself be non-manifold (App. F).

The elimination of indirect dependencies is critical for obtaining a diverse set of triangulations across different resolutions from a single input mesh $\M^L$. 
Each unique vertex split sequence produces a distinct set of intermediate meshes $\{ \hat{\M}^l \}$, which in turn induces a unique split token sequence via $\Sigma_\F$.
Because the choice of vertex split sequences grows combinatorially, this framework facilitates sampling a vast number of diverse upsampling sequences for training a network that generalizes across topologies.

\subsection{Sampling Refinement Sequences} \label{sec:sequence_sampling}
From the space of all possible upsampling sequences, we sample a subset with varied topologies and distinct geometries. 

We obtain diverse topologies by \emph{randomized} Breadth-First Search (BFS) and Depth-First Search (DFS) of the binary forest.
BFS processes all splittable nodes at each tree depth before proceeding deeper, mimicking uniform global refinement. We shuffle same-depth nodes.
DFS fully traverses one subtree before moving to another, mirroring local refinement. 
We randomize root selection and child ordering.
For each encountered node, we obtain geometry and topology tokens.
A key insight is that by progressive mesh construction (\refsec{background}), a decimation sequence defines a single binary forest whose per-node vertex displacements are independent of the upsampling order. Traversals of the same forest thus yield identical geometry tokens and differ only in their topology tokens.

Geometry variation, in contrast, requires perturbing the construction of the binary forest itself: we randomize the Quadric Error edge collapse algorithm \cite{GarlandH97} used to build progressive meshes in \refequ{pm_construction}.
Perturbing vertex quadrics yields different forests, each carrying distinct displacement vectors $\vp_{ki}, \vp_{kj}$, resulting in diversified geometry tokens $\gamma_k$ (\refequ{geometry_token}) for each sequence.
We refer readers to App. B for implementation details.

%% file: text/5_gen_progressive.tex
\section{Learned Adaptive Mesh Refinement}\label{sec:gen_progressive}
The dependency-free tokenizer introduced in \refsec{arbitrary_upsampling} enables extracting diverse upsampling sequences from a single mesh, but it relies on the ground truth binary forest $\F$ to compute vertex split tokens, which is inaccessble during inference. 
In lieu of this, we introduce an autoregressive network that directly predicts tokens using local contexts of a mesh, without requiring the knowledge of the ground truth $\F$. 
We introduce our problem formulation (\refsec{formulation}) and describe the autoregressive model (\refsec{model}).

\begin{figure}[t!]
\centering
\includegraphics[width=\linewidth]{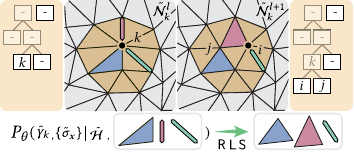}
\caption{
Predicting a vertex split. 
To split $k$, the network predicts token distribution $\mathcal{P}_\theta$ given prior splits $\tilde{\mathcal{H}}$, conditioned on the neighbors (boxed in gray outline).
Geometry tokens $\tilde{\gamma}_k$ displace $k$, while topology tokens $\tilde{\sigma}_x$ $\texttt{R}, \texttt{L}, \texttt{S}$ update neighbors $x \in \tilde{\N}_k^{l+1}$.
The binary forest is then updated (left, right).
}
% \vspace{-10pt}
\Description{}
\label{fig:method_ar_model}
\end{figure}

\begin{figure}[t!]
\centering
\includegraphics[width=\linewidth]{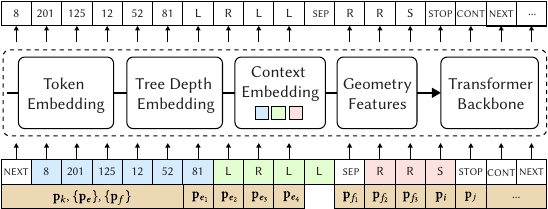}
\caption{
Our autoregressive model. Input tokens (bottom) are first embedded, then added with tree-depth embedding to encode hierarchical structure, and combined with context embeddings to distinguish geometry (blue), edge (green), and face (red) tokens. 
Geometric features (yellow) also condition the transformer backbone to predict the output (top).
}
\Description{}
\label{fig:network}
\end{figure}

%%%%%%%%%%%%%%%%%%%%%%%%%%%%%%%%%%%%%%%%
% Problem Formulation
%%%%%%%%%%%%%%%%%%%%%%%%%%%%%%%%%%%%%%%%

\subsection{Problem Formulation}\label{sec:formulation}
During inference, we initialize a binary forest $\tilde{\F}$ with root nodes corresponding to all vertices on an input mesh. 
We grow this forest by appending child nodes to the parent after each vertex split prediction, contrary to traversal of the ground truth binary forest $\F$ to sample a training sequence.
Given an incomplete, inference binary forest $\tilde{\F}$ after some vertex split predictions, we consider splitting vertex $k$ on mesh $\tilde{\M}$ to vertices $i, j$.
We use $\tilde{\mathcal{H}}$ to denote prediction history and $\tilde{\mathcal{N}}_k \subseteq \tilde{\M}$ to represent the one-ring neighboring simplices around vertex $k$.
The network learns the distribution:
\begin{align}\label{eq:inference}
    \mathcal{P}_{\theta}(\tilde{\gamma}_k, \{ \tilde{\sigma}_x \} \mid \tilde{\mathcal{H}}, \tilde{\N}_k).
\end{align}

Intuitively, conditioned on the previous vertex splits $\tilde{\mathcal{H}}$ and current neighbors $\tilde{\mathcal{N}}_k$, the network learns the joint distribution of the geometry token $\tilde{\gamma}_k$ and the topology tokens $\{ \sigma_x \}$ for all the one-ring simplices $x \in \tilde{\mathcal{N}}_k$.
\reffig{method_ar_model} illustrates this process.
This inference formulation does not require ground truth $\F$, making it suitable for generative tasks. 
It allows the autoregressive model to attend to previously predicted vertex splits, enforcing global structure across refinement steps. 
The conditional input exposes local neighborhood information to the model and enables adaptive inference of specified regions by inputting vertex one-ring geometry features (\refsec{model}).

\begin{figure}[t!]
\centering
\includegraphics[width=0.8\linewidth]{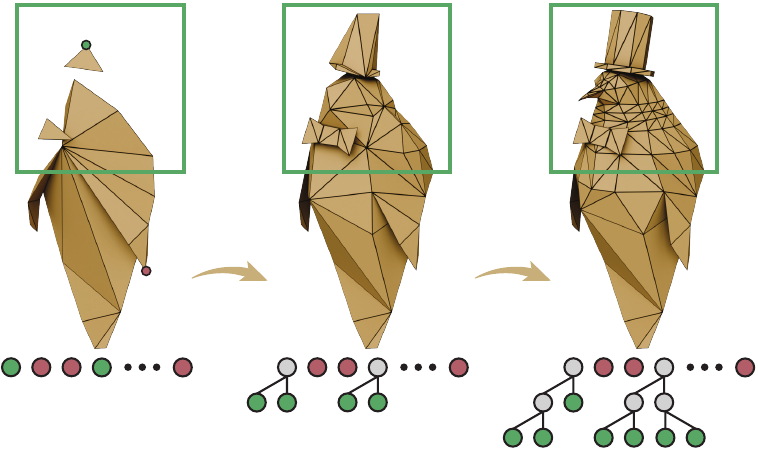}
\caption{
Adaptively refining an input coarse mesh.
Our method directly predicts a binary forest, where active nodes (green circles) that lie within a target region (green box) are fed as input to the autoregressive model to predict child nodes, with vertices outside the target region unchanged (red circles).  
Inactive nodes who have already been split are colored in gray.
}
\Description{}
\label{fig:method_inference}
\end{figure}

%%%%%%%%%%%%%%%%%%%%%%%%%%%%%%%%%%%%%%%%
% Autoregressive Model
%%%%%%%%%%%%%%%%%%%%%%%%%%%%%%%%%%%%%%%%

\subsection{Autoregressive Model} \label{sec:model}

\begin{wrapfigure}[6]{r}{0.75in}
    \vspace{-10pt}
	\includegraphics[width=\linewidth, trim={0.75mm 0mm 0mm 0mm}]{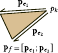}
	\label{fig:method_vectors_e_f} 
\end{wrapfigure}
We propose local geometric conditioning for an autoregressive transformer, using GPT-2 Medium~\cite{taming_transformers, esser2021taming} as the backbone (\reffig{network}).
The token vocabulary consists of geometry tokens (\refsec{geometry_token}), topology tokens (\refsec{vertex_split_tokens}), and control tokens. 
For an upsampling sequence of vertex splits, the tokens for each node are concatenated in split order, where each split follows the pattern $[\texttt{NEXT}, \{\gamma\}, \{\sigma_e\}, \texttt{SEP}, \{\sigma_f\}]$.
We inject local geometric features through input conditioning. 
%
% For a vertex $k$, we project three sets of vectors onto the token embedding space via learnable fully connected layers: vertex positions $\vp_k$, edge vectors $\vp_e$ from
% $k$ to each neighboring vertex, and face spanning vectors $\vp_f$, the concatenation of the two edge vectors spanning each incident face.
We project three sets of vectors onto the token embedding space via learnable fully connected layers: vertex positions $\vp_k$, edge vectors $\vp_e$, and face spanning vectors $\vp_f$.
Each geometry token is conditioned on the full geometric context of the one-ring neighborhood, including the position of the vertex to be split as well as the edge and face vectors of all incident simplices. 
Edge topology tokens $\{\sigma_e\}$ are conditioned on $\vp_e$, and face topology tokens $\{\sigma_f\}$ are conditioned on $\vp_f$. 
Edge and face vectors encode relative positions, adapting naturally to the mesh resolution. They capture global structure at coarse levels and local surface details at fine levels.
Furthermore, we encode the hierarchical structure of the binary forest in the refinement sequence in two ways.
First, we replace sequence-position embedding with tree-depth embedding. Each token's positional embedding is thus determined by the depth of its corresponding node in the binary forest rather than its index in the flattened sequence. 
Second, we use two additional control tokens, \texttt{CONT} and \texttt{STOP}, that signal whether refinement should continue along a given branch. At the end of each vertex split, the model is conditioned on the positions of the newly created vertices $\vp_i, \vp_j$ to predict whether to continue splitting (\texttt{CONT}) or terminate (\texttt{STOP}) each one.
We refer readers to App. A for implementation details.

Given a coarse input mesh, our method refines it by sampling vertices to split. For each candidate vertex $k$, we gather the local geometric features of its one-ring neighborhood ($\vp_k, \vp_e, \vp_f$), pass them in as input conditioning, and the model predicts the corresponding vertex split tokens.
Vertex sampling is guided by the binary forest produced during inference (\refsec{formulation}). At every iteration, every leaf node is a refinement candidate, except for those who has previously received a \texttt{STOP} token. 
Candidates can be visited in either BFS or DFS order, similar to training sequence construction (\refsec{sequence_sampling}). BFS is better suited for global, uniform refinement, while DFS is ideal for local refinement as it descends fully into single subtrees. 
For adaptive local refinement, we both train and run inference with DFS-sampled sequences. Since our architecture employs explicit local conditioning, refining a target region is straightforward and leaves the rest of the mesh untouched: we simply filter the leaf-node candidates to those who fall inside the selected region (\reffig{method_inference}).

%% file: text/6_results.tex
\section{Results and Evaluation} \label{sec:results}

\begin{figure}[t!]
\centering
\includegraphics[width=\linewidth]{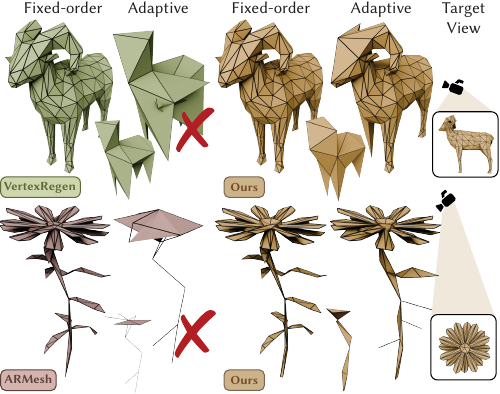}
\caption{
Previous methods, such as VertexRegen \cite{vertexregen} (green) and ARMesh \cite{armesh} (purple), only support training a single fixed-order refinement sequence (left). When the upsampling order changes, both models fail to adapt and cannot be refined any further (right). In contrast, our network is robust to different triangulations and upsampling orders, yielding desired outputs regardless of uniform or adaptive refinement (yellow).
}
\Description{}
\label{fig:results_compare_adap_inf}
\end{figure}

\begin{figure}[t!]
\centering
\includegraphics[width=\linewidth]{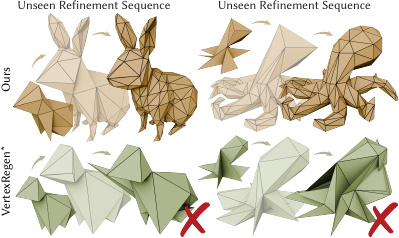}
\caption{
We adapt the tokenization method of VertexRegen~\cite{vertexregen}, denoted VertexRegen*, to train with sequences sampled with our method (\refsec{arbitrary_upsampling}) from a single base mesh. 
Even when trained on 1,000 sequences per shape (versus its original single-sequence supervision), it fails to reconstruct the original mesh given an unseen refinement order (green), while our method succeeds (yellow) on the same data.
}
\Description{}
\label{fig:results_compare_vertex_regen}
\end{figure}

We first evaluate our method against VertexRegen~\cite{vertexregen} and ARMesh~\cite{armesh} on the ShapeNetV2~\cite{shapenet2015} dataset prepared by MeshGPT~\cite{meshgpt}, where we use the chairs and tables categories. 
To match VertexRegen's manifold input requirement, we restrict the training set for all three methods to manifold meshes only, resulting in 660 chairs and 939 tables.
To demonstrate adaptive refinement applications, we use the Toys4K~\cite{toys4k} (1,109 shapes with fewer than 1,200 faces) and Ship-D~\cite{shipd} (2,000 ship hulls with fewer than 800 faces) datasets. 
Toys4K is processed without manifold filtering and includes non-manifold inputs (App. F).
We train a separate network for each dataset.
Additional dataset and training details are provided in App. C and App. D, and we discuss tokenization efficiency and runtime in App. E.

Our experiments demonstrate three key capabilities: 
(1) adaptive refinement of arbitrary vertices, which baseline methods cannot support (\refsec{results_adaptive_refinement}); 
(2) global refinement of coarse geometries (\refsec{results_generalization}); and 
(3) applications (\refsec{results_applications}) such as physics-driven local refinement and coarse-mesh conditioned shape synthesis.

%%%%%%%%%%%%%%%%%%%%%%%%%%%%%%%%%%%%%%%%%%%%
% Adaptive Refinement
%%%%%%%%%%%%%%%%%%%%%%%%%%%%%%%%%%%%%%%%%%%%

\subsection{Adaptive Refinement}
\label{sec:results_adaptive_refinement}

While VertexRegen supports only manifold meshes, ARMesh extends to simplicial complexes.
Both methods tie mesh refinement to a fixed ordering.
Unlike these approaches, our method supports localized upsampling at any vertex in arbitrary refinement orders, enabling view-dependent refinement under face count budget.

We begin with a controlled single-shape comparison: since order-dependence stems from the tokenizer's restriction to a single canonical refinement sequence, this limitation should manifest even when the baselines are trained to memorize a single shape.
When trained on a single upsampling sequence (the standard setting for both baselines), out-of-order refinement causes both to fail (\reffig{results_compare_adap_inf}).
We then test whether this limitation is merely data-driven by training VertexRegen and our method on 1,000 distinct sequences per shape (\reffig{results_compare_vertex_regen}).
Since VertexRegen does not naturally emit multiple sequences, we use \refequ{tracing} to generate diverse upsampling orders and tokenize the resulting vertex splits with their scheme. 
Even with this augmentation, VertexRegen’s fixed-order refinement cannot generalize to unseen refinement sequences, revealing the limitation as architectural rather than data-driven.
In contrast, our dependency-free tokenizer and locally-conditioned model support both fixed-order reconstruction (\reffig{results_compare_adap_inf},~left) and adaptive view-dependent refinement of an input mesh (\reffig{results_compare_adap_inf},~right), where only vertices seen from the viewport are sampled as input to inference (\refsec{model}).

\begin{figure}
\centering
\includegraphics[width=\linewidth]{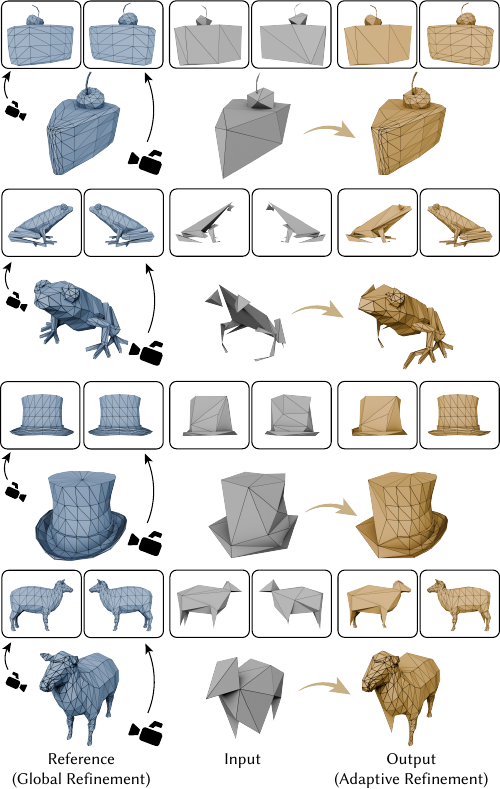}
\caption{
Adaptive inference under face count budget.
Our method, trained on diverse upsampling sequences, allows local inference only on the front view (larger camera) of each object, without upsampling the back (right column, yellow), reducing face count to under 60\% compared to the complete refinement results (left, blue). 
}
\Description{}
\label{fig:results_adaptive_inference}
\end{figure}

Beyond the single-shape evaluations, our method retains adaptive upsampling capability when trained on the Toys4K dataset. We show additional view-dependent refinements where our method reduces face count budget to 60\% of full refinement in \reffig{results_adaptive_inference}.

%%%%%%%%%%%%%%%%%%%%%%%%%%%%%%%%%%%%%%%%%%%%
% Global Refinement
%%%%%%%%%%%%%%%%%%%%%%%%%%%%%%%%%%%%%%%%%%%%

\subsection{Global Refinement}
\label{sec:results_generalization}

\begin{figure}[t!]
\centering
\includegraphics[width=\linewidth]{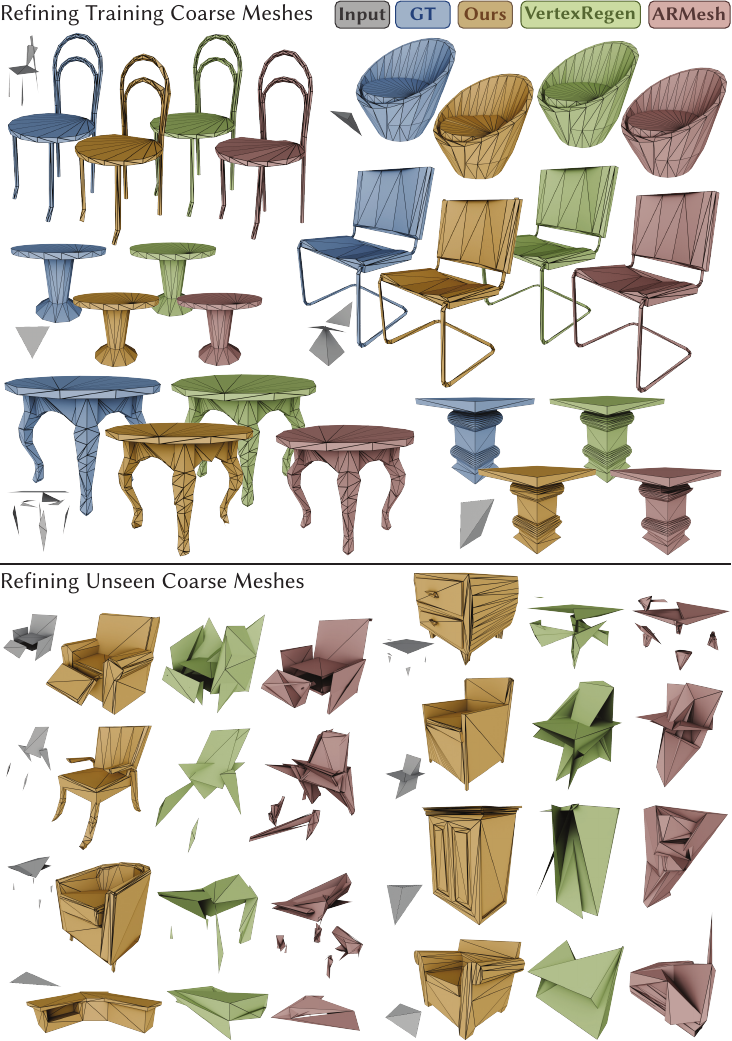}
\caption{
Refining seen (top) and unseen (bottom) coarse meshes (gray). We (yellow) achieve comparable quality as VertexRegen (green) and ARMesh (red) in reconstruction of the ground truth (blue). Our method learns from a rich space of local vertex splits sampled with our tokenizer, enabling refinement of unseen inputs where fixed-order methods fail. 
}
\Description{}
\label{fig:results_recon_refine}
\end{figure}

\begin{table}[t!]
\centering
\setlength{\tabcolsep}{3pt}
\caption{
Quantitative comparisons with baseline methods for refining
unseen coarse meshes of held-out shapes. Best results are in bold.
}
\begin{tabular}{ll|cccc|c}
\multirow{2}{*}{} & \multirow{2}{*}{Method}
 & COV & MMD & 1-NNA & JSD & $\text{EMD}_Q$ \\
 & & ($\uparrow$, \%) & ($\downarrow$, $10^{-3}$) & (50, \%) & ($\downarrow$, $10^{-1}$) & ($\downarrow$, $10^{-2}$) \\
\hline
\multirow{3}{*}{\rotatebox{90}{Chairs}}
 & VertexRegen & \numpct{0.42857142857142855} & \nummmd{0.011861776001751423} & \numpct{0.8571429252624512} & \numjsd{0.4243219265355346} & \numemdq{0.2011443688714856} \\
 & ARMesh      & \numpct{0.38333333333333336} & \nummmd{0.011546839959919453} & \numpct{0.9333333969116211} & \numjsd{0.27831744260721186} & \numemdq{0.3341132691918694} \\
 & Ours        & \bnumpct{0.5423728813559322} & \bnummmd{0.006923966575413942} & \bnumpct{0.5677965879440308} & \bnumjsd{0.16881669083620782} & \bnumemdq{0.014189687730589215} \\
\hline
\multirow{3}{*}{\rotatebox{90}{Tables}}
 & VertexRegen & \bnumpct{0.7058823529411765} & \nummmd{0.007558260578662157} & \numpct{0.29411765933036804} & \numjsd{0.2136751387009626} & \numemdq{0.19840962576776822} \\
 & ARMesh      & \numpct{0.47126436781609193} & \nummmd{0.011002765968441963} & \numpct{0.7988505959510803} & \numjsd{0.22219517831410773} & \numemdq{0.3628153564810486} \\
 & Ours        & \numpct{0.6904761904761905} & \bnummmd{0.004268098622560501} & \bnumpct{0.3154762089252472} & \bnumjsd{0.10751346140741602} & \bnumemdq{0.06490861410274622} \\
\end{tabular}
\label{tab:results_metrics}
\end{table}

\begin{figure*}[t!]
\centering
\includegraphics[width=\linewidth]{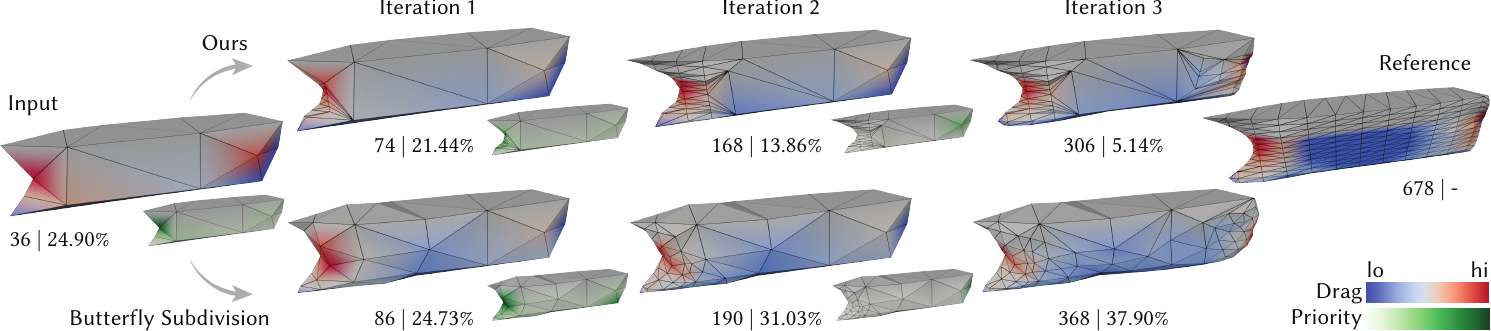}
\caption{
Adaptively refining a coarse mesh given physics-driven signal indicating which regions most affect simulation (App. G), where darker green shows higher influence. 
Our neural method learns the shape prior and refines a coarse mesh to closely match the drag coefficients of the reference hull (red: high drag, blue: low drag).
The face count and relative error to the reference hull's drag are reported with format "count | error".
Classical butterfly subdivision increases triangle count purely based on local geometry without shape semantics, progressively accumulating simulation error.
Our method approximates the reference hull with a tighter face budget, crucial for downstream Boundary Element Method solvers whose evaluation speed depends on face count.
}
\Description{}
\label{fig:results_shiphulls}
\end{figure*}

We evaluate each method's ability to refine coarse meshes from ShapeNetV2~\cite{shapenet2015}.
As a sanity check, all methods achieve comparable reconstruction quality when refining base mesh\-es from the training set (\reffig{results_recon_refine}, top).
To run inference on baselines, we prepend the tokens of the coarsest-level mesh. 
For our method, input mesh vertices are sampled to refine at each step (\refsec{model}).

We further evaluate refinement on coarse meshes that were unseen during training, obtained by different decimations of  training shapes.
This isolates the model's ability to refine \emph{unseen triangulations} of familiar shapes.
Our method produces refinements that add coherent local details, while baselines collapse to incomplete outputs (\reffig{results_recon_refine}, bottom).
We attribute this to our locally-conditioned architecture that learns a broader distribution of valid refinements.

Additionally, we report quantitative results on refining coarse meshes of shapes from the test set. 
Following prior work, we compute point-cloud metrics~\cite{yang2019pointflow} against the test set: Coverage (COV) for diversity, Minimum Matching Distance (MMD) for quality, 1-Nearest Neighbor Accuracy (1-NNA) for distribution similarity, and Jensen-Shannon Divergence (JSD) for spatial similarity.
Our method achieves best results on nearly all metrics (\reftab{results_metrics}, left).
Topologically, both our method and VertexRegen produce manifold outputs, ours via an optional inference-time manifold guarantee (App. F).
We also evaluate triangulation quality with EMD$_Q$, which we define as the earth mover's distance between the distribution of per-triangle mean-ratio quality~\cite{knupp2001algebraic} pooled over all generated meshes and that pooled over the reference meshes.
Our method matches the reference most closely (\reftab{results_metrics}, right).

%%%%%%%%%%%%%%%%%%%%%%%%%%%%%%%%%%%%%%%%%%%%
% Additional Applcations
%%%%%%%%%%%%%%%%%%%%%%%%%%%%%%%%%%%%%%%%%%%%

\subsection{Additional Applications}
\label{sec:results_applications}

\begin{figure}[t!]
\centering
\includegraphics[width=\linewidth]{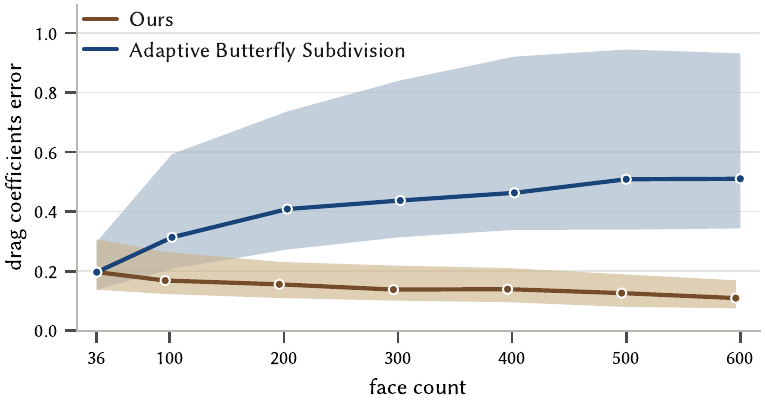}
\caption{
Ship hull drag coefficient error evaluated at different face counts across 100 full-resolution meshes with a median of 657 faces. Solid lines show the median error with the interquartile range shaded. Both methods use the same physics-driven signal that chooses which region to refine (App. G). As face budget grows, our trained model reduces error with learned shape prior while butterfly subdivision increases it by smoothing local geometry. 
}
\Description{}
\label{fig:results_cw_curve}
\end{figure}

\begin{figure}[t!]
\centering
\includegraphics[width=\linewidth]{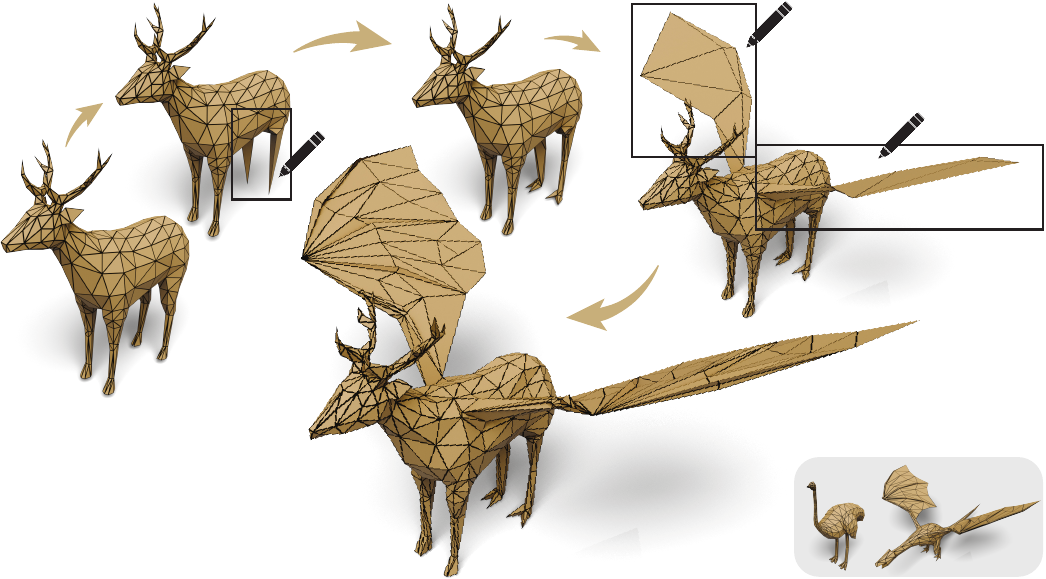}
\caption{
Progressive editing. User coarsely edits an input shape and refines using the trained model to match the mythical creature Peryton (a deer-bird hybrid). Reference shapes are boxed in gray (bottom right).
}
\Description{}
\label{fig:results_editing_deer}
\end{figure}

We present additional applications of our neural method's novel capability of adaptive upsampling by training on the Ship-D dataset for physics-driven local refinement and on the Toys4K dataset for coarse-mesh conditioned shape synthesis.

Better approximation of reference geometry at a tighter face budget is crucial for downstream simulation with Boundary Element Methods such as Capytaine~\cite{ancellin2019capytaine}, whose evaluation speed depends on triangle count. 
To this end, we refine 100 randomly selected coarse ship hulls across a range of face budgets and measure the drag coefficient error against their full-resolution meshes, comparing with adaptive butterfly subdivision.
Both methods perform targeted refinement of regions selected by a physics-driven signal that prioritizes coarse geometry, high curvature, and strong Michell density.
We provide details of this signal and drag coefficients computation in App. G.
We showe an example in \reffig{results_shiphulls} and report quantitative results in \reffig{results_cw_curve}.
Notably, upsampling with the classical subdivision scheme increases error, since it can only refine input triangulation toward a smooth limit surface. 
Our learned method, in contrast, reduces drag coefficient error by refining geometry to match the training data distribution. 
While data-driven baselines also learn shape priors, our dependency-free tokenizer unlocks the ability to refine arbitrary input regions, providing both semantic awareness and adaptive local control. 

\begin{figure}[t!]
\centering
\includegraphics[width=\linewidth]{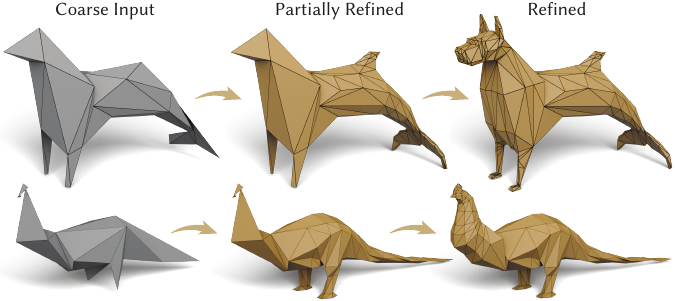}
\caption{
Editing examples of novel synthesized shapes via mix-and-match: a dog-dolphin (top, closest real-world analog is a sea lion) and a chicken-dinosaur (bottom, chickens are descendants of dinosaurs). The user authors a coarse input and uses the trained network to refine the front and back. 
}
\Description{}
\label{fig:results_editing_mix_and_match}
\end{figure}

In mesh editing scenarios, explicit control over which regions to upsample preserves the rest of the input mesh by construction.
A user can make coarse modifications to specific parts of an input mesh, and the model progressively splits vertices within the selected regions.
We show an example of creating a Peryton, a mythical deer–bird hybrid, in \reffig{results_editing_deer}. 
Our method also enables fast prototyping from coarse local edits. 
A user may design a shape from scratch by authoring a coarse mesh, selecting local regions, and using the network to refine them (\reffig{results_editing_mix_and_match}).

%% file: text/7_conclusion.tex
\section{Conclusion and Future Work} \label{sec:conclusion} 
We present a learning-based framework for adaptive mesh refinement that jointly updates local geometry and surface connectivity through vertex split operations. 
Our novel dependency-free tokenizer leverages the progressive mesh forest to sample combinatorially many mesh refinement sequences. 
Our locally-conditioned autoregressive model enables region-selective upsampling and remains robust to unseen coarse triangulations, unlocking applications such as inference-time view-dependent upsampling, physics-driven local refinement, and coarse-mesh conditioned shape synthesis.

Several directions remain for future work. 
Our tokenization treats triangles as the highest-dimensional simplex. Extending to tetrahedra introduces indirect dependencies, as splitting a vertex could create a face outside its one-ring neighborhood. A dependency-free tokenizer for higher dimensions could enable adaptive volumetric modeling with an autoregressive network. 
While we tackle the problem of refining input coarse meshes, our method can be augmented to also generate the coarse mesh tokens using MeshXL’s tokenization scheme~\cite{chen2024meshxl}, similar to VertexRegen~\cite{vertexregen}.
Scaling to larger datasets and broader shape categories is a natural next direction for generalization and novel shape synthesis capabilities.
More broadly, we see learned adaptive refinement as a stepping stone toward 3D pipelines with the granular control of classical tools and the semantic awareness of neural priors.

%% file: text/8_supp.tex
\appendix

%%%%%%%%%%%%%%%%%%%%%%%%%%%%%%%%%%%%%%%
% Autoregressive Model
%%%%%%%%%%%%%%%%%%%%%%%%%%%%%%%%%%%%%%%

\section{Autoregressive Model} \label{app:ar_arch}
We describe our quantization scheme, show the model architecture, and present implementation details.

\subsection{Quantization} \label{app:quantization}
To discretize the displacement vectors $(\mathbf{p}_{ki} = p_i - p_k, \mathbf{p}_{kj} = p_j - p_k)$ for splitting vertex $k$ to $i, j$, we first apply 7-bit quantization to the continuous vertex coordinates of the original mesh. 
The quantized $\mathbf{p}_{ki}, \mathbf{p}_{kj}$ span $[-128, 128]$.
We convert signed integers to unsigned integers and obtain a geometry token $\gamma \in [0, 256]$ per dimension. 
The dependency-free vertex split tokens can be interpreted as classification decisions for each edge and face in the one-ring neighborhood of a vertex, so it's trivial to discretize them.
We exclude the single deterministic $\texttt{LIFT}$ token for vertices.
Additionally, we use \texttt{SOS} as start of sequence, \texttt{NEXT} to denote a new node to split, \texttt{SEP} to separate edge tokens and face tokens, and \texttt{EOS} as end of sequence.
For a node, the list of tokens is $[ \texttt{NEXT}, \{\gamma\}, \{\sigma_e\}, \texttt{SEP}, \{\sigma_f\}]$.
Two new child nodes are created after a node prediction. The network uses two additional control tokens, \texttt{CONT} and \texttt{STOP} to decide whether to continue splitting a child node.
Geometry tokens $\{\gamma\}$ are the concatenation of $x, y, z$ dimensions of $\mathbf{p}_{ki}, \mathbf{p}_{kj}$.
We sort topology tokens $\{\sigma_e\}, \{\sigma_f\}$ by the simplex centroids when sampling training sequences, and inference follows accordingly.
We also use \texttt{PAD} tokens.
In total, we have a vocabulary size of 268.

\subsection{Architecture}
For our transformer backbone, we adopt GPT-2 Medium~\cite{taming_transformers, esser2021taming}, a 350M-parameter architecture. Specifically, we use a hidden dimension of $d=1024$, 24 transformer layers, and 16 attention heads. We use layer normalization inside the residual path, following~\cite{child2019generating, parisotto2020stabilizing} and causal self-attention. Including the learnable parameters detailed below, our architecture has 316M parameters.
We illustrate the architecture in Fig. 9 of the main paper.

\paragraph{Embedding Space}

Each token $t$ is mapped to a $d$-dimensional representation via a learned embedding table $\mathbf{E}_{\text{tok}}$.
To differentiate nodes at different depths in the binary forest, we use learned positional embeddings $\mathbf{E}_{\text{pos}}$ indexed by tree depth.
To differentiate tokens within a node, we use context embeddings $\mathbf{E}_{\text{ctx}}$ that encode three token types: geometry tokens, edge topology tokens and face topology tokens.

\paragraph{Local Geometric Conditioning.}
To condition predictions on local geometry, we project spatial information into the embedding space using learned MLPs.
For vertex position $\vp_k$, we use $\phi_v: \mathbb{R}^3 \to \mathbb{R}^d$. 
For each neighboring edge with direction vector $\vp_e$, we use $\phi_e: \mathbb{R}^3 \to \mathbb{R}^d$.
For each neighboring face with spanning vectors $\vp_f$, we use $\phi_f: \mathbb{R}^6 \to \mathbb{R}^d$.
The networks $\phi_e, \phi_f$ consist of two fully connected layers with a hidden dimension of 1024 and use ReLU activation.
To condition geometry delta tokens, we collect $\{\vp_k, \vp_e, \vp_k\}$ and use a PointNet++~\cite{pointnet++} style permutation-invariant module $\psi$ that maps a variable-size set of neighborhood positions to a fixed-dimensional feature.
Each 3D neighborhood coordinate is passed through a shared 5-layer MLP with a hidden dimension of 1024 and ReLU activation. Afterwards, a 5-layer MLP (1024 dimensions with ReLU) aggregates point features. Features of padded slots are set to $-\infty$ such they are eliminated by the subsequent neighbor-wise max-pool. 
The module $\psi$ that yields a summary of the local geometry.
In all, for geometry tokens $\gamma$, edge topology tokens $\sigma_e$, face topology tokens $\sigma_f$, and child node control tokens, the conditioning signals for a vertex split from $k$ to $i, j$ are:
\begin{equation}
\begin{aligned}
\mathbf{c}^{\gamma}_k &= \psi\big(\vp_k,\; \{\vp_e \mid e \in \mathcal{N}_e(k)\},\; \{\vp_f \mid f \in \mathcal{N}_f(k)\}\big) \\
\mathbf{c}^{\sigma_e}_k &= \phi_e(\vp_e) \\
\mathbf{c}^{\sigma_f}_k &= \phi_f(\vp_f) \\
\mathbf{c}^{v}_i &= \phi_v(\vp_i), \ \mathbf{c}^{v}_j = \phi_v(\vp_j)
\end{aligned}
\label{eq:conditioning}
\end{equation}
where $\mathcal{N}_e(k)$ and $\mathcal{N}_f(k)$ denote the edges and faces in the one-ring neighborhood of $k$.
Additionally, we use the local geometry feature $\mathbf{c}^{\gamma}_k$ to condition the prediction of \texttt{NEXT} and \texttt{EOS} tokens to determine whether the entire refinement is finished. 
The complete input to the transformer backbone for token $t$ at tree depth $\ell$ with context type $n$ is:
\begin{equation*}
    \mathbf{x}_t = \mathbf{E}_{\text{tok}}(t) + \mathbf{E}_{\text{pos}}(\ell) + \mathbf{E}_{\text{ctx}}(n) + \mathbf{c},
\end{equation*}
where we apply the appropriate conditioning $\mathbf{c}$ from \refequ{conditioning} for the specific context type of token $t$.

%%%%%%%%%%%%%%%%%%%%%%%%%%%%%%%%%%%%%%%
% Base Mesh Randomization with Quadric Perturbation
%%%%%%%%%%%%%%%%%%%%%%%%%%%%%%%%%%%%%%%

\section{Base Mesh Randomization with Quadric Perturbation}\label{app:quadric}
When collapsing $i, j$ into $k$, the optimal position $\vp_k = \arg\min_{p} Q(\vp_i) + Q(\vp_j)$ is the minimizer of the summation of vertex quadrics $Q(\vp_i) + Q(\vp_j)$. Our perturbation adds a small noise to the vertex quadrics as $Q'(\vp_v) = Q(\vp_v + \epsilon)$, where $\epsilon \sim \textit{Unif}\,(-\alpha, \alpha)$ is a noise sampled from the uniform distribution with scale $\alpha$. We set $\alpha$ to be $2.5\%$ of the longest mesh bounding box axis. Intuitively, the quadrics determine which edge is cheapest to collapse at each step, so perturbing them changes which vertices get merged together and in what order. The result is a different clustering of the original vertices, and hence a forest with a different hierarchy, rather than the same forest with jittered positions.

%%%%%%%%%%%%%%%%%%%%%%%%%%%%%%%%%%%%%%%
% Dataset
%%%%%%%%%%%%%%%%%%%%%%%%%%%%%%%%%%%%%%%

\section{Dataset Details} \label{app:dataset}
We use ShapeNetV2~\cite{shapenet2015}, Ship-D~\cite{shipd}, and Toys4K~\cite{toys4k} datasets for the experiments presented in the main paper. 
The ShapeNetV2 dataset prepared by MeshGPT \cite{meshgpt} contains meshes under 800 faces. We use the chairs and tables categories and filter for manifold meshes to ensure fair comparison among all three methods (VertexRegen~\cite{vertexregen} that requires manifold inputs, ARMesh~\cite{armesh}, and ours), and we obtain 660 chairs and 939 tables. We use a train/test split of 0.9/0.1. 
Ship-D provides ship hulls generated algorithmically with controllable design features. We use 2,000 hulls sampled from the ``Constrained Randomized Set 2" representing large ships, generated with 13 waterlines along the longitudinal axis and 15 points per waterline, yielding meshes with at most 800 faces.
For the Toys4K dataset, we select meshes under 1,200 faces without manifold constraint, retaining a diverse subset of 1,109 shapes. 

Our model is trained on a collection of pre-generated upsampling sequences. For each shape, we generate 10 distinct base meshes through perturbed Quadric Error decimation (\refapp{quadric}), and from each base mesh we sample 1,000 upsampling sequences via randomized BFS/DFS traversals of the resulting binary forest (Section 5.4). Together, the base-mesh variation exposes the model to diverse coarse inputs, while the per-base sequence variation exposes it to diverse refinement orders. 

To create base meshes, we decimate shapes from all datasets up to 800 vertex splits such that upsampling sequences fit within our context length of 10,624. In practice, this means meshes with fewer than 800 faces are decimated to the minimum (a tetrahedron, or one per component for multi-component meshes), while higher-resolution meshes retain finer base meshes.

The number of upsampling sequences per shape governs how thoroughly the model is exposed to varying refinement orders. 
To select an appropriate value, we conducted a controlled study on 100 ShapeNet chairs in which we fixed one base meshes per shape and varied the number of upsampling sequences per base mesh across \{250, 500, 1000, 2000\}. We evaluated each resulting model by aggregating reconstruction Chamfer distance across multiple inference seeds, which probes the model's ability to refine the same shape under different refinement orderings. 
For each shape, we generate 5 refinement samples per shape under different inference seeds and report the mean Chamfer distance. Chamfer distance ($10^{-3}$) improves from \numfourk{0.008105162023741287} at 250 sequences to \numfourk{0.007111084675416351} at 500 and \numfourk{0.005900668004178442} at 1,000 sequences, with diminishing returns at 2,000 (\numfourk{0.005553356614836957}). We use 1,000 sequences per base mesh in all main experiments, balancing coverage of the refinement-order distribution against the cost of training data generation and storage.

For baselines VertexRegen and ARMesh, we prepare training data with the official codebases.

%%%%%%%%%%%%%%%%%%%%%%%%%%%%%%%%%%%%%%%
% Training
%%%%%%%%%%%%%%%%%%%%%%%%%%%%%%%%%%%%%%%

\section{Training Details} \label{app:training}
We provide training details for experiments in Section 7.

Across all experiments (ShapeNetV2 chairs and tables, Toys4K, and Ship-D), our model uses a maximum context length of 10,624. We apply random translation within range $[-0.1, 0.1]$ as additional data augmentation. We train with the AdamW optimizer using a learning rate of $1 \times 10^{-4}$ with cosine annealing, decreasing gradually to $10^{-6}$. We use mixed-precision training (BF16) with gradient checkpointing and compile the model for additional speedup. Training runs on 8 NVIDIA B200 GPUs with a per-GPU batch size of 32 (effective batch size 256), and takes approximately 2 days for ShapeNetV2 chairs, 3 days for ShapeNetV2 tables, 3 days for Toys4K, and 5 days for Ship-D.

We train VertexRegen~\cite{vertexregen} and ARMesh~\cite{armesh} using the same GPT-2 Medium~\cite{taming_transformers, esser2021taming} backbone as our method, ensuring that architectural capacity is held constant across all comparisons. We adopt each baseline's original tokenization scheme and training protocol, and use the ShapeNetV2 chairs and tables datasets as described in \refapp{dataset} for the experiments in Section 7.2. 
Each baseline is trained on one NVIDIA RTX 5090 GPU until it reaches 100\% training accuracy, ensuring that limitations observed at inference time stem from architectural design rather than insufficient training.
VertexRegen uses an effective batch size of 64 via gradient accumulation and trains for approximately 2 days. Following ARMesh's default token-budgeted batching scheme with a per-GPU budget of 20,000 tokens, it trains for approximately 1 day.

\section{Tokenization Efficiency and Runtime}\label{app:token_runtime}
In terms of tokenization efficiency, the maximum / mean token lengths on ShapeNetV2 chairs and tables are: VertexRegen 4900 / 1052, ARMesh 14225 / 3263, ours 10500 / 3114. VertexRegen is the most compact as it only considers manifold refinement with a maximum of 12 tokens per vertex split, while ARMesh and our method support non-manifold cases, with the token count depending on the number of neighbors for each vertex. 

As for runtime, on the reconstruction task evaluated on the ShapeNetV2 chairs and tables (Section 7.2), the average mesh generation time is: VertexRegen 5.52s, ARMesh 13.14s, ours 14.2s. Our method’s performance is comparable to ARMesh, and VertexRegen’s faster runtime is due to its more compact manifold-only tokenization (see Tokenization Efficiency above). In the view-dependent refinement setting in Fig. 13, our adaptive upsampling not only reduces face budget by 60\% but also runtime by 55\%, a 2.2x speed-up compared to global refinement. Timing results are produced on a machine with a 24GB RTX4090 GPU and a 32GB RAM Ryzen9 CPU.

%%%%%%%%%%%%%%%%%%%%%%%%%%%%%%%%%%%%%%%
% Optional Manifold Guarantees
%%%%%%%%%%%%%%%%%%%%%%%%%%%%%%%%%%%%%%%

\section{Optional Manifold Guarantees}\label{app:manifold}
The design of our network naturally supports non-manifold meshes, giving it the capability to handle a majority of meshes in-the-wild. 
In Fig. 6, we demonstrate an example of non-manifold vertex splits with our tokens.
Results that refine non-manifold input meshes include Fig. 10, the flower in Fig. 11, the frog, hat, and sheep in Fig. 13, and Fig. 18.

For applications that desire manifold outputs, manifoldness can be easily enforced during inference. If splitting a vertex at state $\tilde{\M}^l$ causes the mesh to fail the manifold condition check, we revert to $\tilde{\M}^{l-1}$ and resample another vertex, similar to how the \emph{linked} condition \cite{dey1999topology} is implemented in mesh simplification literature.

%%%%%%%%%%%%%%%%%%%%%%%%%%%%%%%%%%%%%%%
% Physics-Driven Refinement
%%%%%%%%%%%%%%%%%%%%%%%%%%%%%%%%%%%%%%%

\section{Physics-Driven Refinement} \label{app:simulation}

\paragraph{Priority Signal}
We assign a refinement priority $p_v = h_v² \cdot |K_v| \cdot |\partial Y/ \partial x|_v$ for each wetted vertex $v$ below the waterline, where $h_v$ is the mean incident edge length and $|K_v|$ the discrete Gaussian curvature. 
Their product gives a piecewise-linear surface-approximation error indicator. 
$|\partial Y / \partial x|_v$ is the Michell source density on the centerplane, acting as an adjoint weight for the sensitivity of the wave-drag integral to a local surface perturbation. 
At each upsampling step, top 5\% highest-priority vertices are refined, and priorities are recomputed on the resulting mesh.
In this way, refinement concentrates where the mesh is coarse, the hull is curved, and Michell density is most sensitive.

\paragraph{Drag and Error Computation}
Similar to Ship-D~\cite{shipd}, we choose the Michell's Integral~\cite{michell1898xi, tuck1989wave} to obtain drag coefficients at different water depths (25\%, 33\%, 50\%, 67\%) and Froude numbers (0.1 to 0.45 at 0.05 intervals). 
Each mesh therefore yields a 4 by 8 grid of coefficients $C_w$.
%
% We report the mean relative error against the grid of the corresponding full-resolution mesh.
We report the magnitude-weighted relative error
$\sum_{ij} | C_w^{ij} - \hat{C}w^{ij} | / \sum{ij} | \hat{C}_w^{ij} |$
against the grid $\hat{C}_w$ of the corresponding full-resolution mesh.